\documentclass[english,reprint,prd,aps,showpacs]{revtex4-1}

\usepackage[T1]{fontenc}
\usepackage[latin1]{inputenc}
\usepackage[english]{babel}
\usepackage{amsmath}
\usepackage{graphicx}
\usepackage{amssymb}
\usepackage{esint}

\usepackage{longtable}
\usepackage{dcolumn}
\usepackage{babel}

\usepackage{enumerate}
\usepackage{bm}
\usepackage{float}

\makeatother

\makeatletter

\begin{document}

\title{Deformations of the vacuum solutions of general relativity subjected to \\ linear constraints}

\author{C. Molina}
\email{cmolina@usp.br}
\affiliation{Escola de Artes, Ci\^{e}ncias e Humanidades, Universidade de
  S\~{a}o Paulo\\ Av. Arlindo Bettio 1000, CEP 03828-000, S\~{a}o
  Paulo-SP, Brazil}

\begin{abstract}

The problem of deforming geometries is particularly important in the context of constructing new exact solutions of Einstein's equation. This issue often appears when extensions of the general relativity are treated, for instance in brane world scenarios. In this paper we investigate spacetimes in which the energy-momentum tensor obeys a linear constraint. Extensions of the usual vacuum and electrovacuum solutions of general relativity are derived and an exact solution is presented. The classes of geometries obtained include a wide variety of compact objects, among them black holes and wormholes. The general metric derived in this work generalizes several solutions already published in the literature. Perturbations around the exact solution are also considered.

\end{abstract}



\maketitle

\section{Introduction}

Vacuum solutions of the general relativity are of paramount importance. And among those, spherically symmetric solutions are highlighted. Phenomenologically, they describe the gravitational field of approximately spherical astrophysical bodies, while from a more theoretical side they are important systems where ideas involving the quantization of gravity can be implemented.
When the standard four-dimensional Einstein equations are assumed, Birkhoff's theorem and its generalizations fix the spherically symmetric spacetimes as Schwarzschild and its generalizations with non-null cosmological constant $\Lambda$. Considering electrically (or magnetically) charged objects, the electrovacuum solutions are the Reissner-Nordstr\"{o}m metric and its nonasymptotically flat counterparts.

In many scenarios, the relevant question is the issue of extending the general relativity vacuum solutions. That is, the construction of a set of geometries that include the vacuum spacetimes and whose associated energy-momentum tensor satisfies given constraints. 

This issue appears, for example, in the brane world models. In the Shiromizu \textit{et al.} \cite{Shiromizu:1999wj} approach for the Randall-Sundrum scenarios \cite{Randall:1999vf,Randall:1999ee}, black hole and wormhole solutions were constructed \cite{Casadio:2001jg,Bronnikov:2002rn,Bronnikov:2003gx,Molina:2010yu,Molina:2012ay,Neves:2012it} based on the constraint
\begin{equation}
R = 4\Lambda \,\, ,
\label{Ricci_scalar}
\end{equation}
where $R$ is the four-dimensional Ricci scalar. Wormhole geometries were derived \cite{Lobo:2007qi,MontelongoGarcia:2011ag,Wong:2011pt,Cataldo:2008ku} with a constraint in the form
\begin{equation}
\mathcal{P}_{r} = \omega \rho \,\, ,
\label{prho}
\end{equation}
where $\rho$ and $\mathcal{P}_{r}$ are, respectively, the energy density and radial pressure as seen by static observers.
In \cite{Molina:2011mc}, it is presented a large class of isotropic extensions of the vacuum solutions in general relativity, and the relevant constraint was
\begin{equation}
\mathcal{P}_{r} = \mathcal{P}_{t} \,\, ,
\label{isotropic}
\end{equation}
with $\mathcal{P}_{t}$ denoting the tangential pressure. 

In the present work we will focus on a generalization of the relations \eqref{Ricci_scalar}, \eqref{prho} and \eqref{isotropic}. We consider a general linear constraint $\mathcal{L}$ on the components of the energy-momentum tensor associated with spherically symmetric and static geometries. In this setup, our main goal is to obtain extensions of the usual vacuum and electrovacuum solutions of the general relativity. Our results generalize several previous works presented in the literature \cite{Casadio:2001jg,Bronnikov:2002rn,Bronnikov:2003gx,Molina:2010yu,Molina:2011mc,Molina:2012ay,Neves:2012it}.

\section{Linear constraints and exact solutions}

We assume that the field equations for the metric components have the form
\begin{equation}
R_{\nu}^{\mu} - \frac{1}{2} R \, \delta_{\nu}^{\mu} + \Lambda \, \delta_{\nu}^{\mu} = 8\pi T_{\nu}^{\mu} \,\, .
\label{einstein_equations}
\end{equation}
In Eq.~\eqref{einstein_equations}, $T_{\mu\nu}$ is the (at least) effective energy-momentum tensor. Relation \eqref{einstein_equations} naturally includes Einstein's equation, but also some of its generalizations. For instance, setting $8 \pi T_{\mu \nu} = \mathcal{E}_{\mu\nu}$, where $\mathcal{E}_{\mu\nu}$ is the projection of the five-dimensional Weyl tensor on the brane, we obtain the vacuum version of the brane field equations proposed by \cite{Shiromizu:1999wj,Aliev:2005bi}.

In this work we consider spherically symmetric and static geometries. With these assumptions, the metric can be written as 
\begin{equation}
ds^{2} = -A(r) \, dt^{2} + \frac{1}{B(r)} \, dr^{2}
+ r^{2} \, \left( d\theta^{2} + \sin^{2}\theta \, d\phi^{2} \right) \,\, .
\label{spherical_geometry}
\end{equation}

\newpage

\noindent
In the usual coordinate system $(t,r,\theta,\phi)$, the energy-momentum tensor has the form
\begin{equation}
\left[ T_{\nu}^{\mu} \right] =
\left[\begin{array}{cccc}
-\rho &       &      &     \\
      & \mathcal{P}_{r} &      &     \\
      &       & \mathcal{P}_{t} &     \\
      &       &      & \mathcal{P}_{t}
\end{array}
\right] \,\,.
\label{SET-general}
\end{equation}
The energy density ($\rho$) and pressures ($\mathcal{P}_{r}$, $\mathcal{P}_{t}$) defined in Eq.~(\ref{SET-general}) are the quantities seen by static observers, associated to the integral curves of the Killing vector field $\partial/\partial t$.

The basic relation to be explored in this work is a linear constraint $\mathcal{L}$ among $\rho$, $\mathcal{P}_{r}$ and $\mathcal{P}_{t}$, given by
\begin{equation}
\mathcal{L}(\rho,\mathcal{P}_{r},\mathcal{P}_{t}) \equiv \alpha \, \rho + \beta \, \mathcal{P}_{r} + 2 \gamma \, \mathcal{P}_{t} = 0 \,\, .
\label{constraint}
\end{equation}
Within this notation, the brane constraint in Eq.~\eqref{Ricci_scalar} is expressed as $\alpha = -1$, $\beta = 1$ and $\gamma = 1$. The constraint in Eq.~\eqref{prho} is given by $\alpha = \omega$, $\beta = -1$ and $\gamma =0$. And the isotropic condition in Eq.~\eqref{isotropic} reads $\alpha = 0$, $\beta = 1$, $\gamma = -1/2$. Although we keep these cases in mind as references, we are interested in the general linear constraint.
With Eq.~\eqref{constraint} and the field equations~\eqref{einstein_equations}, a differential equation constraining the functions $A$ and $B$ is obtained, 
\begin{gather}
\left( -\alpha + \beta + 2\gamma\right) \Lambda 
+ \frac{\alpha - \beta}{r^{2}}
+ \left( -\alpha + \beta \right) \frac{B(r)}{r^{2}} 
\nonumber \\
- \frac{\gamma}{2} \left[ \frac{A'(r)}{A(r)} \right]^{2}
+ \frac{\gamma}{2} \left[ \frac{A'(r) \, B'(r)}{A(r)} \right] 
+ \frac{\beta + \gamma}{r} \, \left[ \frac{B(r) \, A'(r)}{A(r)} \right]
\nonumber \\
+ \frac{-\alpha + \gamma}{r} \, B'(r)
+ \gamma \, \frac{A''(r) \, B(r)}{A(r)}
= 0 \,\, .
\label{eq_A_B}
\end{gather}
In the present work, (') denotes differentiation with respect to $r$.

We propose to construct a set of solutions assuming that they are ``close'' to the usual spherically symmetric (electro)vacuum solution given by the metric in Eq.~\eqref{spherical_geometry} \linebreak with $B=A=A_{0}$, 
\begin{equation}
A_{0}(r) = 1 - \frac{2M}{r} + \frac{q}{r^{2}} - \frac{\Lambda}{3}r^{2} \,\, ,
\label{vac_solution}
\end{equation}
where $M$ and $q$ are constants. The root structure of $A_{0}$ is an important element in the construction presented here. It is straightforward to check that if $M>0$, and if $q$, $\Lambda$ are smaller than certain values $q_{ext}$, $\Lambda_{ext}$, respectively, then the zeros of $A_{0}$ are simple and at least one of them is real and positive. These conditions will be assumed in this paper.  Here, we are slightly generalizing the Reissner-Nordstr\"{o}m (de Sitter or anti-de Sitter) metric allowing $q$ to be negative.

Denoting a particular solution by the pair $(A,B)$ of functions that satisfies Eq.~\eqref{eq_A_B}, we are searching for a family of solutions $\mathcal{S}$ such that
\begin{enumerate}[(i)]

\item the base solution $(A_{0},A_{0})$ is an element of $\mathcal{S}$;

\item a generic solution $(A_{\alpha},B_{\alpha})\in\mathcal{S}$ is
a continuous deformation of $(A_{0},A_{0})$; that is, there is (at
least) one set of solutions $\mathcal{D} = \left\{ \left(A_{C},B_{C}\right), C_{min} \leq C \leq C_{max} \right\} $, labeled by a real parameter $C$, such that $\mathcal{D} \subset \mathcal{S}$.
\end{enumerate}
Since we are interested in deformations of the vacuum solutions, we assume the \emph{ansatz}
\begin{equation}
A(r) = A_{0}(r) \,\, ,
\label{ansatz_A}
\end{equation}
\begin{equation}
B(r) = A_{0}(r) + \left(C - 1\right) \, B_{lin}(r) \,\, .
\label{ansatz_B}
\end{equation}
A specific choice for the normalization of the constant $C$ is a matter of convenience. It is immediate from Eq.~\eqref{ansatz_B} that if $C=1$, the vacuum solution is recovered. Therefore, according to condition (i), the constant $C$ must assume values in a subset of $\mathbb{R}$ that includes $C=1$. 
In Sec.~\ref{perturbation}, this \emph{ansatz} will be further generalized.

The exact solution for the correction $B_{lin}$, general up to a multiplicative integration constant, is given by
\begin{equation}
B_{lin}(r) = \exp \left\{ - \int \frac{f(r)}{A_{0}(r) \, h(r)} \, dr   \right\}  \,\, ,
\label{b_linear}
\end{equation}
with the functions $h$ and $f$ defined as
\begin{equation}
h(r) = 
\left( \gamma - \alpha \right) \, A_{0}(r) + \frac{\gamma}{2} \, r \, A_{0}'(r)
 \,\, ,
\label{function_h}
\end{equation}
\begin{gather}
f(r) = 
(\beta - \alpha) \frac{A_{0}(r)}{r}
+ (\alpha - \beta - 2\gamma) \Lambda \, r A_{0}(r) 
\nonumber \\
+ (\alpha - \gamma) \, A_{0}(r) \, A_{0}'(r)
- \frac{\gamma}{2} \, r \left[ A_{0}'(r) \right]^{2}
 \,\, .
\label{function_f}
\end{gather}
Because $A_{0}$ in Eq.~\eqref{vac_solution} is a rational function, it follows that $h$, $f$ and the integrand in Eq.~\eqref{b_linear} are also rational functions. Expanding this integrand in partial fractions, an explicit expression for $B$ is readily obtained for any linear constraint $\mathcal{L}$.

\section{General characteristics of the deformed solutions}

We notice that not all constraints are compatible with arbitrary values of $q$. In fact, if $\gamma =0$, the admissible base solution $(A_{0},A_{0})$ is only admissible if $q=0$ in Eq.~\eqref{vac_solution}, that is, taking as the base solution the Schwarzschild (S), Schwarzschild-de Sitter (S-dS) or  Schwarzschild-anti-de Sitter (S-AdS) metric; if the cosmological constant is null, positive or negative respectively. 
On the other hand, if $\gamma \ne 0$, the constraint parameters must satisfy the relation $\alpha - \beta + 2\gamma = 0$ so that $q\ne0$, and therefore the Reissner-Nordstr\"{o}m (RN),  Reissner-Nordstr\"{o}m-de Sitter (RN-dS) or Reissner-Nordstr\"{o}m-anti-de Sitter (RN-AdS) metric is included in the family of extensions constructed. We summarize these results in Table~\ref{relations}.
\begin{table}[H]

\caption{Relations between constraints and admissible base solutions.}
\label{relations}

\begin{tabular*}{\columnwidth}{*{4}{c@{\extracolsep{\fill}}}}
\hline
\hline
& \textbf{Constraint class} & \textbf{Admissible base solutions} & \\ 
\hline
& $\gamma = 0$  &  
S, S-dS, S-AdS & 
\\
\hline
& $\gamma \ne 0$ and $\alpha - \beta + 2\gamma \ne 0$ & 
S, S-dS, S-AdS &
\\
\hline
& $\gamma \ne 0$ and $\alpha - \beta + 2\gamma = 0$ & 
S, S-dS, S-AdS and
&
\\
& & RN, RN-dS, RN-AdS & \\
\hline
\hline

\end{tabular*}
\end{table}

Once the solutions given by Eqs.~\eqref{ansatz_A} and \eqref{ansatz_B} are explicitly derived, one important question is the determination of the submanifold in which static observers are present, that is, where the integral curves of $\partial/ \partial t$ are timelike. This region is determined by $r\in I$, where $I=\{r\in\mathbb{R}^{+}, A(r)>0, B(r)>0\}$, and into it the coordinate system $(t,r,\theta,\phi)$ is well defined. 
With this static submanifold settled, the next step is the construction of its maximal extension. Some important loci are
\begin{itemize}

\item \textbf{Killing horizon at $\bm{r=r_{+}}$.} \\
We denote $r_{+}$ as the largest zero of $A_{0}$ with $A_{0}'(r_{+})>0$. The surface $r=r_{+}$ is a candidate to a Killing horizon (and event horizon).
This follows from the fact that $h(r_{+}) = \gamma A_{0}'(r_{+})/2 \ne 0$ and $f(r_{+}) = - \gamma \left[ A_{0}'(r_{+})\right]^{2}/2 \ne 0$. Therefore
\begin{equation}
A(r) \sim B(r) \sim  \left( r - r_{+} \right)
\,\, ,
\end{equation}
which implies that $r=r_{+}$ is a Killing horizon.

\item \textbf{Killing horizon at $\bm{r=r_{c}}$.} \\
We denote $r_{c}$ as a zero of $A_{0}$ with $r_{c}>r_{+}$. If $\Lambda>0$, the function $A_{0}$ has at least two positive zeros $r_{+}$ and $r_{c}$, such that $r_{c}>r_{+}$ and $A(r)>0$ for $r\in(r_{+},r_{c})$. In this case, $A'(r_{+})>0$ and $A'(r_{c})<0$. The surface $r=r_{c}$ is a candidate to a Killing horizon--a cosmological horizon. The argument is analogous to the one presented for the $r=r_{+}$ case.

\item \textbf{Curvature singularity at $\bm{r=r_{0}}$.} \\
We denote $r_{0}$ as the largest zero of $h$. It is straightforward to show that $h$ has at least one real positive zero. The limit $r \rightarrow r_{0}$ is a candidate to a curvature singularity. In fact, we have $h'(r_{0})\ne 0$ and therefore
\begin{equation}
B_{lin}(r) \sim \frac{1}{\left( r - r_{0} \right)^{ c_{0}}}
\,\, ,
\end{equation}
with
\begin{equation}
c_{0} = \lim_{r\rightarrow r_{0}} \frac{f(r)}{A_{0}(r) h'(r)} \,\, .
\label{c0_geral}
\end{equation}
When $c_{0}>0$ and $r_{0}\in I$ (for example considering the constraints discussed in \cite{Casadio:2001jg,Bronnikov:2002rn,Bronnikov:2003gx,Molina:2010yu,Molina:2012ay}), the function $B$ and the scalar curvatures diverge in the limit $r\rightarrow r_{0}$. Since $r_{0} \in I$ for a wide range of the parameters that characterize a given geometry, for these cases a curvature singularity is present.

\item \textbf{Timelike trapping horizon at $\bm{r=r_{thr}}$.} \\
We denote $r_{thr}$ a simple zero of $B$, but not of $A_{0}$. If $C<1$, it is possible that $A_{0}(r_{thr})\ne 0$ with $A'_{0}(r_{thr})\ne 0$ and $B(r_{thr})=0$. The timelike surface $r=r_{thr}$ is a candidate to a trapping horizon when it is part of the maximal extension.

\end{itemize}

If $C>0$, the Killing horizon $r=r_{+}$ (and $r=r_{c}$, when $r_{c}$ exists) is nonextreme, with a non-null surface gravity. The maximal extension of the geometry in this case is made with the standard techniques, for example expressing the metric in terms of the null coordinates $u=t-r_{\star}$ or $v=t+r_{\star}$, where $r_{\star}$ is the ``tortoise coordinate'',
\begin{equation}
\frac{dr_{\star}(r)}{dr} = \frac{1}{\sqrt{A(r)B(r)}} \,\, .
\label{tartaruga}
\end{equation}
The spacetime in this case describes a black hole with a non-null surface gravity.

If $C=0$ and a static region is present, the Killing horizon $r=r_{+}$ will be extreme. The solution models a black hole with a null surface gravity. This is possible even with $q<q_{ext}$ and $\Lambda<\Lambda_{ext}$, as we are assuming. The extension in this case can be made, for example, using the ``quasi-global'' radial coordinate $w$ \cite{Bronnikov:2003gx,Bronnikov:2008by}, defined as
\begin{equation}
\frac{dw(r)}{dr} = \sqrt{\frac{A(r)}{B(r)}} \,\, .
\label{quasi_global}
\end{equation}

If $C<0$ and there is a trapping horizon at $r=r_{thr}$, the extension can be made expressing the metric in terms of the tortoise coordinate $r_{\star}$, defined in Eq.~\eqref{tartaruga}. The geometry is extended to $\{(t,r_{\star},\theta,\phi), -\infty < r_{\star} < \infty \}$. This spacetime describes a wormholelike structure with a throat at $r_{\star}=0$ ($r=r_{thr}$).

\section{Beyond the exact solution}
\label{perturbation}

We now generalize the \emph{ansatz} in Eqs.~\eqref{ansatz_A} and \eqref{ansatz_B}. We assume that the components of the energy-momentum tensor in Eq.~\eqref{SET-general} are smooth functions of a parameter $\delta$ such that with $\delta=0$ and $C=1$, we recover the vacuum solutions. In this way, the energy density ($\rho$) and pressures ($\mathcal{P}_{r}$, $\mathcal{P}_{t}$) can be written as
\begin{equation}
\rho = \rho_{0} + \left( C - 1 \right) \rho^{lin}  + \sum_{n=1} \delta^{n} \, \rho_{(n)} \, \, ,
\label{rho-expansion}
\end{equation}
\begin{equation}
\mathcal{P}_{r} = \mathcal{P}_{r0} + \left( C - 1 \right) \mathcal{P}_{r}^{lin} + \sum_{n=1} \delta^{n} \, \mathcal{P}_{r\,(n)} \, \, ,
\label{pr-expansion}
\end{equation}
\begin{equation}
\mathcal{P}_{t} = \mathcal{P}_{t0} + \left( C - 1 \right) \mathcal{P}_{t}^{lin} + \sum_{n=1} \delta^{n} \, \mathcal{P}_{t\,(n)} \, \, .
\label{pt-expansion}
\end{equation}
The terms $\rho_{0}$, $\mathcal{P}_{r 0}$ and $\mathcal{P}_{t 0}$ are the energy density and pressures associated with the vacuum solution. 

Given the degree of generality adopted, a simple form for the constraint in the metric components is not available. We propose a perturbative approach, considering the limit of small $\delta$. We have effectively two deformation parameters:  $C$, assuming values in a subset of $\mathbb{R}$;  and $\delta$, so that $|\delta|\ll 1$ and the perturbation around the exact solution derived in the previous section is meaningful.

Considering the specific form of the solutions already derived, we write the metric functions, up to the first order in $\delta$, as
\begin{equation}
A(r) = A_{0}(r) \left[ 1 + \delta a_{1}(r) \right]  + \mathcal{O} \left(\delta^{2}\right) \,\, ,
\label{general_A_nl}
\end{equation}
\begin{equation}
B(r) = A_{0}(r) \left[ 1 +  \left( C - 1 \right) \, b_{lin}(r) + \delta b_{1}(r) \right] + \mathcal{O} \left(\delta^{2}\right)\,\, ,
\label{general_B_nl}
\end{equation}
where  $b_{lin}$ is defined as
\begin{equation}
B_{lin}(r) =  A_{0}(r) \, b_{lin}(r) \,\, .
\label{blin}
\end{equation}
We obtain finally a simple linear relation between the corrections $b_{1}$, $b_{1}'$, $a_{1}'$, $a_{1}''$:
\begin{equation}
P_{1}(r) \, b_{1}'(r) + P_{2}(r) \, b_{1}(r) + P_{3}(r) \, a_{1}''(r) + P_{4}(r) \, a_{1}'(r) = 0 \,\, ,
\label{eq_vinculo_linear}
\end{equation}
with the functions $P_{1}$, $P_{2}$, $P_{3}$ e $P_{4}$ given by
\begin{eqnarray}
P_{1}(r) & = &  
2\left(\gamma - \alpha\right) \, r \, A_{0}(r) + \gamma r^{2} \, A_{0}'(r) \,\, ,
\label{P1} \\
P_{2}(r) & = & 
2\left[
\left(\beta - \alpha\right) \, A_{0}(r)
+ \left(2\gamma + \beta - \alpha\right) \, r \, A_{0}'(r) 
 \right. \nonumber \\
& & \left. + \gamma r^{2} A_{0}''
\right] \,\, , 
\label{P2} \\
P_{3}(r) & = & 
2 \gamma r^{2}  \left[1 + \left(C - 1\right) b_{lin}(r) \right]\, A_{0}(r)  \,\, ,
\label{P3} \\
P_{4}(r) & = & 
3 \gamma r^{2} \left[1 + \left(C - 1\right) b_{lin}(r) \right] \, A_{0}'(r)
\nonumber \\
& & + r \left\{ 
2 \left( \beta + \gamma \right) \left[ 1 + \left(C - 1\right) b_{lin}(r) \right] \right. 
\nonumber \\
& &  + \left. \gamma \left( C - 1 \right) r \, b_{lin}'(r)
\right\} \, A_{0}(r) \,\, .
\label{P4}
\end{eqnarray}

\section{Concluding remarks}

We have presented an explicit procedure to construct extensions of the spherically symmetric and static vacuum solutions of general relativity. The extensions are characterized by a linear constraint in components of the energy-momentum tensor. The general metric derived in this work generalizes several solutions already published in the literature. The new solutions describe a wide variety of compact objects, including singular and nonsingular black holes, wormholes and naked singularities.

Particular solutions of the presented work have already been extensively explored in the literature \cite{Casadio:2001jg,Bronnikov:2002rn,Bronnikov:2003gx,Molina:2010yu,Molina:2012ay,Molina:2011mc}. Still, the possibilities of the formalism presented here for linear constraints are broader than the particular cases already considered. For instance, a constraint equation in the form $\mathcal{P}_{r}= \omega \,\rho$ admits a deceptive simple solution for the metric,
\begin{eqnarray}
ds^{2} & = & - \left( 1 - \frac{2M}{r} \right) dt^{2} 
%
+  \left[  1 - \frac{2M}{r} + \frac{C-1}{r \left( r - 2M \right)^{1/\omega} } \right] dr^{2}
\nonumber \\
& & + r^{2} d\Omega^{2}_{2} .
\label{new_solution}
\end{eqnarray}
The geometry described by Eq.~\eqref{new_solution} remains to be fully understood. In fact, a large set of spacetimes can be obtained for the different values of $\omega$, including black holes \cite{Molina:2011mc}.
 
Going beyond the general solution constructed, we show the form of the relation among the metric components, assuming a perturbative approach. Given the linearity of the relation~\eqref{eq_vinculo_linear}, the effect of several applied perturbations can be independently considered, with the compound energy-momentum tensor given by the sum of its several components. This point makes the developed formalism very flexible, as seen previously in more particular contexts \cite{Molina:2011mc}.

\vspace{0.5cm}

\begin{acknowledgments}
This work was partially supported by Conselho Nacional de Desenvolvimento Cient\'{\i}fico e Tecnol\'ogico (CNPq), Grant No.~303431/2012-1; and Funda\c{c}\~{a}o de Amparo \`{a} Pesquisa do Estado de S\~{a}o Paulo (FAPESP), Grant No.~2012/15775-5; Brazil.
\end{acknowledgments}


\begin{thebibliography}{10}%
\makeatletter
\providecommand \@ifxundefined [1]{%
 \ifx #1\undefined \expandafter \@firstoftwo
 \else \expandafter \@secondoftwo
\fi
}%
\providecommand \@ifnum [1]{%
 \ifnum #1\expandafter \@firstoftwo
 \else \expandafter \@secondoftwo
\fi
}%
\providecommand \enquote [1]{``#1''}%
\providecommand \bibnamefont  [1]{#1}%
\providecommand \bibfnamefont [1]{#1}%
\providecommand \citenamefont [1]{#1}%
\providecommand\href[0]{\@sanitize\@href}%
\providecommand\@href[1]{\endgroup\@@startlink{#1}\endgroup\@@href}%
\providecommand\@@href[1]{#1\@@endlink}%
\providecommand \@sanitize [0]{\begingroup\catcode`\&12\catcode`\#12\relax}%
\@ifxundefined \pdfoutput {\@firstoftwo}{%
 \@ifnum{\z@=\pdfoutput}{\@firstoftwo}{\@secondoftwo}%
}{%
 \providecommand\@@startlink[1]{\leavevmode}%
 \providecommand\@@endlink[0]{}%
}{%
 \providecommand\@@startlink[1]{%
  \leavevmode
  \pdfstartlink
   attr{/Border[0 0 1 ]/H/I/C[0 1 1]}%
   user{/Subtype/Link/A<</Type/Action/S/URI/URI(#1)>>}%
  \relax
 }%
 \providecommand\@@endlink[0]{\pdfendlink}%
}%
\providecommand \url  [0]{\begingroup\@sanitize \@url }%
\providecommand \@url [1]{\endgroup\@href {#1}{\urlprefix}}%
\providecommand \urlprefix [0]{URL }%
\providecommand \Eprint[0]{\href }%
\@ifxundefined \urlstyle {%
  \providecommand \doi [1]{doi:\discretionary{}{}{}#1}%
}{%
  \providecommand \doi [0]{doi:\discretionary{}{}{}\begingroup
  \urlstyle{rm}\Url }%
}%
\providecommand \doibase [0]{http://dx.doi.org/}%
\providecommand \Doi[1]{\href{\doibase#1}}%
\providecommand \bibAnnote [3]{%
  \BibitemShut{#1}%
  \begin{quotation}\noindent
    \textsc{Key:}\ #2\\\textsc{Annotation:}\ #3%
  \end{quotation}%
}%
\providecommand \bibAnnoteFile [2]{%
  \IfFileExists{#2}{\bibAnnote {#1} {#2} {\input{#2}}}{}%
}%
\providecommand \typeout [0]{\immediate \write \m@ne }%
\providecommand \selectlanguage [0]{\@gobble}%
\providecommand \bibinfo [0]{\@secondoftwo}%
\providecommand \bibfield [0]{\@secondoftwo}%
\providecommand \translation [1]{[#1]}%
\providecommand \BibitemOpen[0]{}%
\providecommand \bibitemStop [0]{}%
\providecommand \bibitemNoStop [0]{.\EOS\space}%
\providecommand \EOS [0]{\spacefactor3000\relax}%
\providecommand \BibitemShut [1]{\csname bibitem#1\endcsname}%
\bibitem{Shiromizu:1999wj}%
  \BibitemOpen
  \bibfield{author}{%
  \bibinfo {author} {\bibfnamefont{T.}~\bibnamefont{Shiromizu}}, \bibinfo
  {author} {\bibfnamefont{K.-i.}\ \bibnamefont{Maeda}},\ and\ \bibinfo {author}
  {\bibfnamefont{M.}~\bibnamefont{Sasaki}},\ }%
  \bibfield{journal}{%
  \Doi{10.1103/PhysRevD.62.024012}{\bibinfo {journal} {Phys. Rev. D}}\ }%
  \textbf{\bibinfo {volume} {62}},\ \bibinfo {pages} {024012} (\bibinfo {year}
  {2000}),\ \Eprint{http://arxiv.org/abs/gr-qc/9910076}{arXiv:gr-qc/9910076}%
  \bibAnnoteFile{NoStop}{Shiromizu:1999wj}%
\bibitem{Randall:1999vf}%
  \BibitemOpen
  \bibfield{author}{%
  \bibinfo {author} {\bibfnamefont{L.}~\bibnamefont{Randall}}\ and\ \bibinfo
  {author} {\bibfnamefont{R.}~\bibnamefont{Sundrum}},\ }%
  \bibfield{journal}{%
  \Doi{10.1103/PhysRevLett.83.4690}{\bibinfo {journal} {Phys. Rev. Lett.}}\ }%
  \textbf{\bibinfo {volume} {83}},\ \bibinfo {pages} {4690} (\bibinfo {year}
  {1999}),\ \Eprint{http://arxiv.org/abs/hep-th/9906064}{arXiv:hep-th/9906064}%
  \bibAnnoteFile{NoStop}{Randall:1999vf}%
\bibitem{Randall:1999ee}%
  \BibitemOpen
  \bibfield{author}{%
  \bibinfo {author} {\bibfnamefont{L.}~\bibnamefont{Randall}}\ and\ \bibinfo
  {author} {\bibfnamefont{R.}~\bibnamefont{Sundrum}},\ }%
  \bibfield{journal}{%
  \Doi{10.1103/PhysRevLett.83.3370}{\bibinfo {journal} {Phys. Rev. Lett.}}\ }%
  \textbf{\bibinfo {volume} {83}},\ \bibinfo {pages} {3370} (\bibinfo {year}
  {1999}),\ 
  \Eprint{http://arxiv.org/abs/hep-ph/9905221}{arXiv:hep-ph/9905221}%
  \bibAnnoteFile{NoStop}{Randall:1999ee}%
\bibitem{Casadio:2001jg}%
  \BibitemOpen
  \bibfield{author}{%
  \bibinfo {author} {\bibfnamefont{R.}~\bibnamefont{Casadio}}, \bibinfo
  {author} {\bibfnamefont{A.}~\bibnamefont{Fabbri}},\ and\ \bibinfo {author}
  {\bibfnamefont{L.}~\bibnamefont{Mazzacurati}},\ }%
  \bibfield{journal}{%
  \Doi{10.1103/PhysRevD.65.084040}{\bibinfo {journal} {Phys. Rev. D}}\ }%
  \textbf{\bibinfo {volume} {65}},\ \bibinfo {pages} {084040} (\bibinfo {year}
  {2002}),\ \Eprint{http://arxiv.org/abs/gr-qc/0111072}{arXiv:gr-qc/0111072}%
  \bibAnnoteFile{NoStop}{Casadio:2001jg}%
\bibitem{Bronnikov:2002rn}%
  \BibitemOpen
  \bibfield{author}{%
  \bibinfo {author} {\bibfnamefont{K.~A.}\ \bibnamefont{Bronnikov}}\ and\
  \bibinfo {author} {\bibfnamefont{S.-W.}\ \bibnamefont{Kim}},\ }%
  \bibfield{journal}{%
  \Doi{10.1103/PhysRevD.67.064027}{\bibinfo {journal} {Phys. Rev. D}}\ }%
  \textbf{\bibinfo {volume} {67}},\ \bibinfo {pages} {064027} (\bibinfo {year}
  {2003}),\ \Eprint{http://arxiv.org/abs/gr-qc/0212112}{arXiv:gr-qc/0212112}%
  \bibAnnoteFile{NoStop}{Bronnikov:2002rn}%
\bibitem{Bronnikov:2003gx}%
  \BibitemOpen
  \bibfield{author}{%
  \bibinfo {author} {\bibfnamefont{K.~A.}\ \bibnamefont{Bronnikov}}, \bibinfo
  {author} {\bibfnamefont{V.~N.}\ \bibnamefont{Melnikov}},\ and\ \bibinfo
  {author} {\bibfnamefont{H.}~\bibnamefont{Dehnen}},\ }%
  \bibfield{journal}{%
  \Doi{10.1103/PhysRevD.68.024025}{\bibinfo {journal} {Phys. Rev. D}}\ }%
  \textbf{\bibinfo {volume} {68}},\ \bibinfo {pages} {024025} (\bibinfo {year}
  {2003}),\ \Eprint{http://arxiv.org/abs/gr-qc/0304068}{arXiv:gr-qc/0304068}%
  \bibAnnoteFile{NoStop}{Bronnikov:2003gx}%
\bibitem{Molina:2010yu}%
  \BibitemOpen
  \bibfield{author}{%
  \bibinfo {author} {\bibfnamefont{C.}~\bibnamefont{Molina}}\ and\ \bibinfo
  {author} {\bibfnamefont{J.~C.~S.}\ \bibnamefont{Neves}},\ }%
  \bibfield{journal}{%
  \Doi{10.1103/PhysRevD.82.044029}{\bibinfo {journal} {Phys. Rev. D}}\ }%
  \textbf{\bibinfo {volume} {82}},\ \bibinfo {pages} {044029} (\bibinfo {year}
  {2010}),\ \Eprint{http://arxiv.org/abs/1005.1319}{arXiv:1005.1319}%
  \bibAnnoteFile{NoStop}{Molina:2010yu}%
\bibitem{Molina:2012ay}%
  \BibitemOpen
  \bibfield{author}{%
  \bibinfo {author} {\bibfnamefont{C.}~\bibnamefont{Molina}}\ and\ \bibinfo
  {author} {\bibfnamefont{J.~C.~S.}\ \bibnamefont{Neves}},\ }%
  \bibfield{journal}{%
  \Doi{10.1103/PhysRevD.86.024015}{\bibinfo {journal} {Phys. Rev. D}}\ }%
  \textbf{\bibinfo {volume} {86}},\ \bibinfo {pages} {024015} (\bibinfo {year}
  {2012}),\ \Eprint{http://arxiv.org/abs/1204.1291}{arXiv:1204.1291}%
  \bibAnnoteFile{NoStop}{Molina:2012ay}%
\bibitem{Neves:2012it}%
  \BibitemOpen
  \bibfield{author}{%
  \bibinfo {author} {\bibfnamefont{J.~C.~S.}\ \bibnamefont{Neves}}\ and\
  \bibinfo {author} {\bibfnamefont{C.}~\bibnamefont{Molina}},\ }%
  \bibfield{journal}{%
  \Doi{10.1103/PhysRevD.86.124047}{\bibinfo {journal} {Phys. Rev. D}}\ }%
  \textbf{\bibinfo {volume} {86}},\ \bibinfo {pages} {124047} (\bibinfo {year}
  {2012}),\ \Eprint{http://arxiv.org/abs/1211.2848}{arXiv:1211.2848}%
  \bibAnnoteFile{NoStop}{Neves:2012it}%
\bibitem{Lobo:2007qi}%
  \BibitemOpen
  \bibfield{author}{%
  \bibinfo {author} {\bibfnamefont{F.~S.~N.}\ \bibnamefont{Lobo}},\ }%
  \bibfield{journal}{%
  \Doi{10.1103/PhysRevD.75.064027}{\bibinfo {journal} {Phys. Rev. D}}\ }%
  \textbf{\bibinfo {volume} {75}},\ \bibinfo {pages} {064027} (\bibinfo {year}
  {2007}),\ \Eprint{http://arxiv.org/abs/gr-qc/0701133}{arXiv:gr-qc/0701133}%
  \bibAnnoteFile{NoStop}{Lobo:2007qi}%
\bibitem{MontelongoGarcia:2011ag}%
  \BibitemOpen
  \bibfield{author}{%
  \bibinfo {author} {\bibfnamefont{N.}~\bibnamefont{Montelongo~Garcia}}\ and\
  \bibinfo {author} {\bibfnamefont{F.~S.}\ \bibnamefont{Lobo}},\ }%
  \bibfield{journal}{%
  \Doi{10.1142/S021773231103739X}{\bibinfo {journal} {Mod. Phys. Lett. A}}\ }%
  \textbf{\bibinfo {volume} {26}},\ \bibinfo {pages} {3067} (\bibinfo {year}
  {2011}),\ \Eprint{http://arxiv.org/abs/1106.3216}{arXiv:1106.3216}%
  \bibAnnoteFile{NoStop}{MontelongoGarcia:2011ag}%
\bibitem{Wong:2011pt}%
  \BibitemOpen
  \bibfield{author}{%
  \bibinfo {author} {\bibfnamefont{K.}~\bibnamefont{Wong}}, \bibinfo {author}
  {\bibfnamefont{T.}~\bibnamefont{Harko}},\ and\ \bibinfo {author}
  {\bibfnamefont{K.}~\bibnamefont{Cheng}},\ }%
  \bibfield{journal}{%
  \Doi{10.1088/0264-9381/28/14/145023}{\bibinfo {journal} {Classical Quantum
Gravity}}\
  }%
  \textbf{\bibinfo {volume} {28}},\ \bibinfo {pages} {145023} (\bibinfo {year}
  {2011}),\ \Eprint{http://arxiv.org/abs/1105.2605}{arXiv:1105.2605}%
  \bibAnnoteFile{NoStop}{Wong:2011pt}%
\bibitem{Cataldo:2008ku}%
  \BibitemOpen
  \bibfield{author}{%
  \bibinfo {author} {\bibfnamefont{M.}~\bibnamefont{Cataldo}}, \bibinfo
  {author} {\bibfnamefont{S.}~\bibnamefont{del Campo}}, \bibinfo {author}
  {\bibfnamefont{P.}~\bibnamefont{Minning}},\ and\ \bibinfo {author}
  {\bibfnamefont{P.}~\bibnamefont{Salgado}},\ }%
  \bibfield{journal}{%
  \Doi{10.1103/PhysRevD.79.024005}{\bibinfo {journal} {Phys. Rev. D}}\ }%
  \textbf{\bibinfo {volume} {79}},\ \bibinfo {pages} {024005} (\bibinfo {year}
  {2009}),\ \Eprint{http://arxiv.org/abs/0812.4436}{arXiv:0812.4436}%
  \bibAnnoteFile{NoStop}{Cataldo:2008ku}%
\bibitem{Molina:2011mc}%
  \BibitemOpen
  \bibfield{author}{%
  \bibinfo {author} {\bibfnamefont{C.}~\bibnamefont{Molina}}, \bibinfo {author}
  {\bibfnamefont{P.}~\bibnamefont{Martin-Moruno}},\ and\ \bibinfo {author}
  {\bibfnamefont{P.~F.}\ \bibnamefont{Gonzalez-Diaz}},\ }%
  \bibfield{journal}{%
  \Doi{10.1103/PhysRevD.84.104013}{\bibinfo {journal} {Phys. Rev. D}}\ }%
  \textbf{\bibinfo {volume} {84}},\ \bibinfo {pages} {104013} (\bibinfo {year}
  {2011}),\ \Eprint{http://arxiv.org/abs/1107.4627}{arXiv:1107.4627}%
  \bibAnnoteFile{NoStop}{Molina:2011mc}%
\bibitem{Aliev:2005bi}%
  \BibitemOpen
  \bibfield{author}{%
  \bibinfo {author} {\bibfnamefont{A.~N.}\ \bibnamefont{Aliev}}\ and\ \bibinfo
  {author} {\bibfnamefont{A.~E.}\ \bibnamefont{Gumrukcuoglu}},\ }%
  \bibfield{journal}{%
  \Doi{10.1103/PhysRevD.71.104027}{\bibinfo {journal} {Phys. Rev. D}}\ }%
  \textbf{\bibinfo {volume} {71}},\ \bibinfo {pages} {104027} (\bibinfo {year}
  {2005}),\ \Eprint{http://arxiv.org/abs/hep-th/0502223}{arXiv:hep-th/0502223}%
  \bibAnnoteFile{NoStop}{Aliev:2005bi}%
\bibitem{Bronnikov:2008by}%
  \BibitemOpen
  \bibfield{author}{%
  \bibinfo {author} {\bibfnamefont{K.~A.}\ \bibnamefont{Bronnikov}}, \bibinfo
  {author} {\bibfnamefont{E.}~\bibnamefont{Elizalde}}, \bibinfo {author}
  {\bibfnamefont{S.~D.}\ \bibnamefont{Odintsov}},\ and\ \bibinfo {author}
  {\bibfnamefont{O.~B.}\ \bibnamefont{Zaslavskii}},\ }%
  \bibfield{journal}{%
  \Doi{10.1103/PhysRevD.78.064049}{\bibinfo {journal} {Phys. Rev. D}}\ }%
  \textbf{\bibinfo {volume} {78}},\ \bibinfo {pages} {064049} (\bibinfo {year}
  {2008}),\ \Eprint{http://arxiv.org/abs/0805.1095}{arXiv:0805.1095}%
  \bibAnnoteFile{NoStop}{Bronnikov:2008by}%
\end{thebibliography}

%

\end{document}